\documentclass[11pt]{article}
\newlength{\vshift}
\newlength{\hshift}
\usepackage{amssymb,amsopn}

\def\la{\lambda}
\def\La{\Lambda}

\def\de{\delta}
\def\be{\beta}
\def\al{\alpha}

\def\shouldid{\stackrel{!}{=}}

\def\ds{\stackrel{\star}{,}}

\def\p{\partial}

\def\hp{\hat{\partial}}
\def\h{\hat}
\def\lb{\lbrack}
\def\rb{\rbrack}
\def\pat{\partial}

\begin{document}

\begin{titlepage}
\rightline{LMU-TPW 2003-10}
\rightline{MPP-2003-106}

\vspace{4em}
\begin{center}

{\Large{\bf Gauge theories on the $\kappa$-Minkowski  spacetime}}

\vskip 3em

{{\bf Marija Dimitrijevi\' c${}^{1,2,3}$,
    Frank Meyer${}^{1,2}$, \\  Lutz M\"oller${}^{1,2}$, Julius Wess${}^{1,2}$ }}

\vskip 1em

${}^{1}$Universit\"at M\"unchen, Fakult\"at f\"ur Physik\\
        Theresienstr.\ 37, D-80333 M\"unchen\\[1em]

${}^{2}$Max-Planck-Institut f\"ur Physik\\
        F\"ohringer Ring 6, D-80805 M\"unchen\\[1em]

${}^{3}$University of Belgrade, Faculty of Physics\\
Studentski trg 12, SR-11000 Beograd\\[1em]
 \end{center}

\vspace{2em}

\begin{abstract}
This study of gauge field theories on $\kappa$-deformed Minkowski spacetime extends
previous work on field theories on this example of a noncommutative
spacetime. We construct deformed gauge theories for arbitrary compact
Lie groups using the concept of enveloping algebra-valued gauge
transformations and the Seiberg-Witten formalism. Derivative-valued
gauge fields lead to field strength tensors as the sum of curvature-
and torsion-like terms. We construct the Lagrangians
explicitly to first
order in the deformation parameter. This is the first
example of a gauge theory that possesses a deformed Lorentz
covariance.
\end{abstract}
\vskip 1.5cm
\qquad\hspace{2mm}\scriptsize{eMail: dmarija;  meyerf;  lmoeller;
 wess@theorie.physik.uni-muenchen.de}
\vfill

\end{titlepage}\vskip.2cm

\newpage
\setcounter{page}{1}

\section*{Introduction}

The best known description of the fundamental forces of nature is given
by gauge theories. Nevertheless intrinsic difficulties arise in these
theories at very high energies or very short distances. Physics is not
very well known in this limit. This has lead to the idea of modifying
the structure of spacetime at very short distances and to introduce
uncertainty relations for the coordinates to provide a natural
cut-off (for reviews of this wide field see \cite{nekrasov}, \cite{szabo}). It is expected that gauge theories still play a vital role in
this regime.

We expect especially interesting new features of gauge field theories
formulated on spaces with a deformed spacetime symmetry. Here we
concentrate on the $\kappa$-deformed Poincar\'e algebra (introduced in
\cite{lukrue1}, \cite{lukrue2} and \cite{majrue}\footnote{For
  additional references concerning this model see \cite{f1}.}). The spacetime
which is covariant with respect to this deformed symmetry algebra is
called the $\kappa$-deformed quantum space.

In a previous paper \cite{f1} deformed field theories
on a $\kappa$-deformed quantum space have been
constructed. The techniques necessary for such
a construction have been thoroughly discussed there.
In this paper we show how the deformation concept
can be applied to a gauge
field theory. We construct deformed gauge theories
 for arbitrary compact Lie groups. ``Deformed'' does not mean that the Lie
groups will be deformed, however, the transformation
parameters will depend on the elements
of the $\kappa$-deformed coordinate space. This implies that
Lie algebra  gauge transformations are generalized to enveloping algebra-valued gauge
transformations.

This is possible by making use of the Seiberg-Witten map \cite{SW},
\cite{madore}, \cite{jurco}. This
is a map that allows to express all elements of the noncommutative
gauge theory by their commutative analogs. It follows that a deformed gauge theory can
be constructed with exactly the same number of fields
(gauge fields, matter fields) as the standard gauge
theory on undeformed space.

Of special interest is the interplay of the gauge
transformations with the $\kappa$-deformed Lorentz
transformations. Gauge theories are based on the concept of covariant
derivatives constructed with gauge fields. Covariance
now refers both to the gauge transformations and to the
$\kappa$-Lorentz transformations.

Theories like the one presented here can be used to deform the
Standard Model (compare with the approach in \cite{moeller} and \cite{wohlgenannt}).
For example, new coupling terms in the Lagrangian arise. This has experimental
consequences and the model can be tested phenomenologically.
We exhibit these terms for an arbitrary
gauge group to first order in the deformation parameter.
These models should be understood in such an expansion,
it renders an infrared cutoff. We do not assume that these
models should be used to describe physics at large
distances.

To obtain phenomenologically interesting results we develop
the theory on a spacelike $\kappa$-deformed spacetime with Minkowski signature
(in \cite{f1} $\kappa$-deformed
Euclidean spacetime was discussed).

The paper is organized as follows: In the first section we present
a compilation of all relevant formulae for $\kappa$-Minkowski
spacetime. To derive and understand these formulae, \cite{f1} is
essential.
In the second section we investigate covariant
derivatives for enveloping algebra-valued gauge transformations.
Attention is given to the $\kappa$-Lorentz covariance as well
as to gauge covariance. For this purpose the enveloping
algebra-valued gauge formalism is developed and the
transformation property of the gauge field is
derived. This leads to the new concept of derivative-valued gauge fields. The field strength tensors, defined
by commutators of covariant derivatives, are
derivative-dependent as well. We expand them  in terms of covariant
derivatives and show that each expansion coefficient is a tensor
by itself; we call them torsion-like tensors. The
derivative-independent term is a deformation of $F^0_{\mu \nu}$
and is used in the construction of Lagrangians.

In the third section we construct the Seiberg-Witten map.
We use the $\star$-product formalism and expand in the
deformation parameter. All gauge and matter fields of the deformed
theory can  be expressed in terms of the standard Lie algebra gauge
fields and the standard matter fields. This
allows the construction of a Lagrangian in terms of the
standard fields.

In the fourth section we discuss the interplay of $\kappa$-Lorentz
and gauge transformations. We show that gauge transformations
and $\kappa$-Lorentz transformations commute and that the
$\kappa$-Lorentz transformed gauge transformation reproduces again
the algebra. This can be implemented in a more abstract setting,
but we discuss this issue explicitely in order to become familiar
with the comultiplication rules and their consequences.

\section{The $\kappa$-Minkowski space}

In a previous paper we discussed the $\kappa$-Euclidean
space \cite{f1} (introduced in \cite{lukrue1} and \cite{lukrue2}) and argued that the generalization to
a Minkowski version is straightforward. We introduce here this
spacelike $\kappa$-deformed Minkowski spacetime, which
is more interesting for phenomenological applications.
First we present the relevant
formulae.

\vspace*{0.2cm}
\noindent
{\bf Coordinate space}

\vspace*{0.1cm}

We start from the same algebra as in \cite{f1}:
\begin{equation}
\label{k1}
[\hat{x}^\mu,\hat{x}^\nu]=i(a^\mu\hat{x}^\nu-a^\nu\hat{x}^\mu), \qquad \mu,\nu=0,1,\dots,n,
\end{equation}
but the metric $\eta^{\mu\nu}=diag(1, -1, \dots, -1)$ and its inverse
are used to raise
and lower indices.
Spacelike deformation will be achieved by
assuming $a^\mu$ to be spacelike. The $n$-direction
is rotated in the direction of $a^\mu$, $a^n=a$, $a^j=0$.
We label  the $n$ commuting coordinates
with $\hat{x}^i$ ($i=0,\dots, n-1$) as opposed to $\hat{x}^n$ and obtain the
following commutation relations:
\begin{equation}
\label{k3}
[\hat{x}^n,\hat{x}^j]=ia\hat{x}^j,\quad [\hat{x}^i,\hat{x}^j]=0,
\quad i,j=0,1,\dots,n-1.
\end{equation}
The parameter $a$ is related to the frequently used parameter
$\kappa$:
\begin{equation}
\label{k4}
a={1\over \kappa}.
\end{equation}

\vspace*{0.2cm}
\noindent
{\bf The $\kappa$-deformed Lorentz algebra}

\vspace*{0.1cm}

The formulae for the transformation of the spacetime
coordinates
are the same as for the Euclidean space,
replacing $\delta^{\mu\nu}$ with  $\eta^{\mu\nu}$ and paying attention to upper and
lower indices:
\begin{eqnarray}
\label{k5}
\lb M^{ij},\hat x^\mu\rb &=& \eta^{\mu j} \hat x^i - \eta^{\mu i} \hat x^j ,
\nonumber\\
\lb M^{in}, \hat x^\mu\rb &=& \eta^{\mu n} \hat x^i - \eta^{\mu i} \hat x^n + iaM^{i\mu},
\end{eqnarray}
$\mu =0,1,\dots n.$ These relations are consistent with the algebra
(\ref{k3}) and they lead to the undeformed algebra relations
of the generators $M^{\mu\nu}$ of the Lorentz algebra $so(1,n)$:
\begin{equation}
\label{k2}
\lb M^{\mu\nu},M^{\rho\sigma}\rb = \eta^{\mu\sigma}M^{\nu\rho} + \eta^{\nu\rho} M^{\mu\sigma} - \eta^{\mu\rho}M^{\nu\sigma}
- \eta^{\nu\sigma} M^{\mu\rho}.
\end{equation}As in \cite{f1}, this is again the undeformed
algebra \cite{majrue}. However, the generators act in a deformed way on products of
functions (i.e. they have a deformed coproduct)
\begin{eqnarray}
\label{k5a}
M^{ij}(\hat{f}\cdot\hat{g}) &=& (M^{ij}\hat{f})\cdot \hat{g}+\hat{f}\cdot (M^{ij}\hat{g}) ,\nonumber\\
M^{in}(\hat{f}\cdot\hat{g}) &=&  (M^{in}\hat{f})\cdot \hat{g}+(e^{ia\hp_n}\hat{f})\cdot (M^{in}\hat{g})
+ia(\hp_k\hat{f})\cdot(M^{ik}\hat{g}).
\end{eqnarray}
In this paper we adopt the convention that over double Latin indices
should be summed from $0$ to $n-1$ and over double Greek from $0$ to $n$.

\vspace*{0.2cm}
\noindent
{\bf Derivatives}

\vspace*{0.1cm}

We introduce derivatives such that they are consistent with (\ref{k3}):
\begin{eqnarray}
\label{k6}
\lb \hp _n , \hat x^\mu\rb  &=& \eta_n^{\ \mu}, \nonumber\\
\lb\hp _i , \hat x^\mu \rb &=& \eta_i^{\ \mu} -ia\eta^{\mu n}\hp _i .
\end{eqnarray}
Derivatives naturally carry a lower index.
It is possible to derive from  (\ref{k6}) the Leibniz rule (i.e. the coproduct):
\begin{eqnarray}
\label{k7}
\hat{\partial}_n\hspace{1mm} (\hat{f} \cdot \hat{g}) & = & (\hat{\partial}_n \hat{f})
\cdot \hat{g} + \hat{f} \cdot (\hat{\partial}_n \hat{g}),\nonumber \\
\hat{\partial}_i\hspace{1mm} (\hat{f} \cdot \hat{g}) & = & (\hat{\partial}_i \hat{f})
\cdot \hat{g} + (e^{ia\hat{\partial}_n} \hat{f}) \cdot (\hat{\partial}_i \hat{g}).
\end{eqnarray}
The derivatives are a $\kappa$-Lorentz algebra
module:
\begin{eqnarray}
\label{k8}
\lb M^{ij}, \hp _\mu\rb &=& \eta ^j_{\ \mu} \hp ^i -  \eta ^i_{\ \mu} \hp ^j, \nonumber\\
\lb M^{in}, \hp _n\rb &=& \hp ^i, \\
\lb M^{in}, \hp _j\rb &=& \eta^i_{\ j}{e^{2ia\hp _n}-1\over 2ia}  -
{ia\over 2}\eta^ i_{\ j} \hp^l\hp _l+ ia\hp
^i\hp _j .\nonumber
\end{eqnarray}

The part of $M^{\mu \nu}$ that acts on the coordinates
and derivatives
can be expressed in terms of the coordinates and the derivatives:
\begin{eqnarray}
\label{k9}
\hat{M}^{ij} &=&\hat{x}^i\hat{\partial}^j -\hat{x}^j\hat{\partial}^i,\nonumber\\
\hat{M}^{in} &=&\hat{x}^i{1-e^{2ia\hat{\partial}_n}\over 2ia}-\hat{x}^n\hat{\partial}^i
+{ia\over 2}\hat{x}^i\hp^l\hp _l.
\end{eqnarray}

\vspace*{0.2cm}\newpage
\noindent
{\bf Dirac operator}

The $\kappa$-deformed Dirac operator has the components:
\begin{eqnarray}
\hat{D}_n &=& {1\over a} \sin (a\hp _n)-
{ia\over 2}\hp^l\hp _le^{-ia\hp _n}, \nonumber \\
\hat{D}_i &=&\hp _i e^{-ia\hp _n}, \label{k11}\\
\lb M^{\mu\nu}, \hat{D}_\rho\rb &=& \eta^{\nu}_{\ \rho}\hat{D}^\mu- \eta^{\mu}_{\ \rho}\hat{D}^\nu.
\end{eqnarray}
It can be seen as a derivative as well and satisfies
the Leibniz rule:
\begin{eqnarray}
\label{k12}
\hat{D}_n\hspace{1mm} (\hat{f} \cdot \hat{g}) & = & (\hat{D}_n \hat{f})
\cdot (e^{-ia\hat{\partial}_n}\hat{g}) + (e^{ia\hat{\partial}_n}\hat{f}) \cdot (\hat{D}_n \hat{g})-ia(\hat{D}_ie^{ia\hat{\partial}_n}\hat{f})\cdot (\hat{D}^i \hat{g}),\nonumber \\
\hat{D}_i\hspace{1mm} (\hat{f} \cdot \hat{g}) & = & (\hat{D}_i \hat{f})
\cdot (e^{-ia\hat{\partial}_n}\hat{g}) +  \hat{f} \cdot (\hat{D}_i \hat{g}). 
\end{eqnarray}
That the Dirac operator really acts as a derivative follows from the commutation relations, when we expand the square root:
\begin{eqnarray}
\label{k15}
\lb \h{D}_n, \hat{x}^j\rb  &=& -ia\h{D}^j , \nonumber \\
\lb \h{D}_n, \hat{x}^n\rb &=& \sqrt{1+a^2{\hat D}^\mu{\hat D}_\mu} , \nonumber \\
\lb \h{D}_i, \hat{x}^j\rb &=& \eta _i^{\ j}\left(
-ia{\hat D}_n+\sqrt{1+a^2{\hat D}^\mu{\hat D}_\mu}\right) ,\\
\lb \h{D}_i, \hat{x}^n\rb &=& 0.\nonumber
\end{eqnarray}

\vspace*{0.2cm}
\noindent
{\bf $\star$-product}
\vspace*{0.1cm}

In the $\star$-product formulation (explained in detail in \cite{f1}) all the elements of the coordinate
algebra can be realized as functions of commuting variables. Derivatives
and the $\kappa$-Lorentz algebra generators can be realized in terms of  commuting
variables and derivatives. On the $\star$-product of functions they act with their
comultiplication rules without seeing the $x$ and derivative dependence
of the $\star$-product.

Here we present for convenience a compilation of the relevant formulae used in the rest of
this paper.

 The $\kappa$-Minkowski spacetime (\ref{k3}) can
be considered as a Lie algebra with
$C^{\mu\nu}_\la=a(\eta^\mu_{\ n}\eta^\nu_{\ \la}-\eta^\nu_{\ n}\eta^\mu_{\ \la})$ as structure constants.
These structure constants appear also in the expansion of the symmetric $\star$-product:
\begin{footnotesize}\begin{eqnarray}
\label{s3}
f \star  g\hspace{1mm} (x)& = &\lim_{{y \rightarrow x \atop z \rightarrow x}}
\exp\Big(\frac{i}{2}x^\la C^{\mu\nu}_\la\p_\mu\otimes\p_\nu 
- \frac{a}{12}x^\la C^{\mu\nu}_\la (\partial_n\partial_\mu \otimes \partial_\nu
- \partial_\mu \otimes \partial_n\partial_\nu)+\dots
\Big)\hspace{1mm} f (y) \otimes g (z)
\nonumber\\
&=& f(x) g(x) + \frac{ia}{2} x^j (\p_n f(x) \p_j g(x) -
\p_jf(x) \p_n  g(x))+ \dots  .
\end{eqnarray}\end{footnotesize}
The derivatives
\begin{eqnarray}
\p ^*_n f(x) & = & \p _n f(x), \nonumber\\
\p _i^* f(x) & = & {e^{ia\p _n} -1\over ia\p _n}\p _i  \, f(x) \label{k16}
\end{eqnarray}
have the Leibniz rule
\begin{eqnarray}\label{k17}
\pat_n^* ( f(x)\star g(x) ) &=&\left(\pat_n^* f(x)\right) \star g(x)
+ f(x)\star \left( \pat_n^* g(x) \right),\nonumber\\
\pat^*_i ( f(x)\star g(x) ) &=& \left( \pat_i^* f(x) \right) \star g(x)
 + ( e^{ia \pat_n^*} f(x) ) \star \left( \pat_i^* g(x)\right).
\end{eqnarray}
The Dirac operator
\begin{eqnarray}
D^*_n f(x) &=& \left( {1\over a} \sin (a\p _n) -
        {\cos (a\pat_n) - 1\over ia\pat_n^2}\p _j\p ^j \right) f(x),\nonumber\\
D^*_i f(x) &=& {e^{-ia\p _n}-1 \over -ia\p _n}\p _i f(x)  \label{k18}
\end{eqnarray}
has the following Leibniz rule
\begin{eqnarray}
\label{k19}
D^*_n ( f(x)\star g(x) ) &=& (D_n^* f(x)) \star
        (e^{-ia \pat_n^*} g(x)) \nonumber + (e^{ia\pat_n^*} f(x)) \star
        (D_n^* g(x))\nonumber\\
&& - ia\left( D_j^* e^{ia\pat_n^*} f(x) \right) \star
        ({D^j}^* g(x)), \nonumber\\
D_i^* (  f(x)\star g(x) ) &=& ( D_i^* f(x)) \star
        (e^{-ia\pat_n^*} g(x))  + f(x) \star (D_i^* g(x)).\nonumber
\end{eqnarray}
The generators of $\kappa$-Lorentz transformations,
acting on coordinates and derivatives
\begin{eqnarray} \label{k20}
M^{*in} f(x) & = & \Big( x^i\p ^n - x^n\p ^i+ x^i\p _\mu\p^ \mu {e^{ia \p _n}-1\over 2\p _n}
 + x^\mu\p _\mu\p ^i {a\p _n+i(e^{ia\p _n}-1)\over a\p _n^2} \Big) f(x),\nonumber\\
M^{*ij} f(x) & = & \left(x^i\p ^j-x^j\p ^i \right) f(x), 
\end{eqnarray}
have the following coproduct
\begin{eqnarray}
 \label{k21}
M^{*in} \left( f(x)\star g(x)\right) &=& \left( M^{*in} f(x) \right) \star g(x)
 + \left(e^{ia \pat_n^*} f(x)\right) \star \left( M^{*in} g(x)\right)\nonumber\\
&& + ia \left( \pat_j^* f(x) \right) \star \left( M^{*ij} g(x)\right), \\
M^{*ij} \left(  f(x)\star g(x)\right)&=&( M^{*ij}   f(x)) \star g(x)+ f(x) \star ( M^{*ij}   g(x)).
\end{eqnarray}
Thus, the entire calculus developed in the abstract algebra can be formulated in
the $\star$-product setting. For applications in physics this
is of advantage because functions of commuting variables
$x$ are suitable representations of physical objects like fields.

\vspace*{0.3cm}
\section{Covariant Derivatives}

Gauge theories will be formulated with the help of covariant
derivatives. We shall demand covariance under the $\kappa$-Lorentz
algebra as well as covariance under the gauge group.
Gauge fields have to be vector fields
that transform like the Dirac operator under the deformed
Lorentz algebra to render a covariant theory.

\vspace*{0.2cm}
\noindent
{\bf $\kappa$-Lorentz covariance}

\vspace*{0.1cm}
We start from the definition of a scalar field. In an undeformed
theory this would be $\phi'(x')=\phi(x)$.
For noncommuting variables we try however to avoid $\hat{x}'$.
Note that
\begin{equation}
\label{g1a}
\hat{x}'^\mu \hat{x}'^\nu \neq
(1+\epsilon_{\mu\nu}\hat{M}^{\mu\nu})\hat{x}^\mu \hat{x}^\nu.\nonumber
\end{equation}
Therefore we replace $\hat{\phi}'(\hat{x}')$ with
$(1+\epsilon_{\mu\nu}\hat{M}^{\mu\nu})\phi'(\hat{x})$, where $\hat{M}^{\mu\nu}$ acts on the coordinates and the derivatives and has been
defined in (\ref{k9}).
The defining equation for a scalar field will take the form
\begin{equation}
\label{g2a}
(1+\epsilon_{\mu\nu}\hat{M}^{\mu\nu})\hat{\phi}'(\hat{x})=\hat{\phi}(\hat{x}),
\end{equation}
with the immediate consequence
\begin{equation}
\label{g3a}
\hat{\phi}'(\hat{x})=\hat{\phi}(\hat{x})-\epsilon_{\mu\nu}\hat{M}^{\mu\nu}\hat{\phi}(\hat{x}).
\end{equation}

To compute the transformation law of a derivative of a scalar field we
calculate $(1+\epsilon_{\mu\nu}\hat{M}^{\mu\nu})\hat{D}_\rho\hat{\phi}'(\hat{x})$
that replaces $\hat{D}'_\rho\hat{\phi}'(\hat{x}')$:
\begin{small}
\begin{eqnarray}
\label{g4a}
(1+\epsilon_{\mu\nu}\hat{M}^{\mu\nu})\hat{D}_\rho\hat{\phi}'(\hat{x})
&=&\hat{D}_\rho(1+\epsilon_{\mu\nu}\hat{M}^{\mu\nu})\hat{\phi}'(\hat{x}) +\epsilon_{\mu\nu}\lb
\hat{M}^{\mu\nu},\hat{D}_\rho\rb\hat{\phi}'(\hat{x}) \nonumber\\
&=&\hat{D}_\rho\hat{\phi}(\hat{x})+\epsilon_{\mu\nu}(\eta^\nu_{\ \rho}\hat{D}^\mu
-\eta^\mu_{\ \rho}\hat{D}^\nu)\hat{\phi}(\hat{x}).
\end{eqnarray}
\end{small}
We have used (\ref{g2a}) to obtain this result and the fact that the
second part on the right hand side is already $\epsilon$-linear.

The transformation law of a derivative of a scalar field is used
to define the transformation law of a vector field:
\begin{equation}
\label{g5a}
(1+\epsilon_{\mu\nu}\hat{M}^{\mu\nu})\hat{V}'_\rho(\hat{x})
=\hat{V}_\rho(\hat{x})+\epsilon_{\mu\nu}(\eta^\nu_{\ \rho}\hat{V}^\mu-\eta^\mu_{\ \rho}\hat{V}^\nu).
\end{equation}
Thus, the derivative
\begin{equation}
\label{g6a}
\hat{\mathcal{D}}_\rho=\hat{D}_\rho-i\hat{V}_\rho
\end{equation}
is $\kappa$-Lorentz covariant.

\vspace*{0.2cm}
\noindent
{\bf Gauge covariant derivatives}

\vspace*{0.1cm}
Gauge theories are based on a gauge group. This
is a compact Lie group with generators $T^a$:
\begin{equation}
\label{g3}
[T^a,T^b]=if^{ab}_{\hspace{3mm} c} T^c.
\end{equation}
Fields are supposed to span linear representations
of this group. Infinitesimal transformations
with constant parameters $\alpha_a$
are:
\begin{equation}
\label{g4}
\de_\al \psi = i\al \psi, \qquad \al := \sum_a \al_aT^{a}=\al_aT^{a}.
\end{equation}
As usual, $\alpha$ is Lie algebra-valued and the
commutator of two such transformations closes
into a Lie algebra-valued transformation:
\begin{equation}
\label{g5}
(\de_\al \de_\be - \de_\be \de_\al )\psi  = [\al,\be]\psi=
i\al_a\be_b f^{ab}_{\hspace{3mm} c}T^c\psi
\equiv \de_{\al\times\be}\psi .
\end{equation}
The symbol $\alpha \times \beta$ is defined by this
equation and it is independent of the
representation of the generators $T^a$.

We generalize the gauge transformations (\ref{g4}) by
considering $\hat{x}$-dependent parameters
$\hat{\alpha}_a(\hat{x})$. Whereas for commuting coordinates Lie algebra-valued
transformations close in the Lie algebra, this will
not be true for noncommuting coordinates. This effect of
noncommutativity leads to a generalization of Lie algebra-valued gauge
transformations \cite{madore}, \cite{jurco}.

There are exactly two representation-independent
concepts based on the commutation relations (\ref{g3}). These
 are the Lie algebra and the enveloping algebra. The
enveloping algebra of the Lie algebra is defined as the
free algebra generated by the elements $T^a$ and
then divided by the ideal generated by the commutation
relations (\ref{g3}). It is infinite-dimensional and
consists of all the (abstract) products of the generators
modulo the relations (\ref{g3})\footnote{Note: the product is not the
matrix product of the generators in a particular
representation.}. Two elements of the enveloping algebra are
identified if they can be transformed into each other
by the use of the commutation relations (\ref{g3}).

A basis can be chosen for the enveloping algebra, we use the symmetrized
products as such a basis and denote
elements of the basis with colons:
\begin{eqnarray}
\label{g6}
:T^a:&=&T^a, \nonumber\\
:T^aT^b:&=&\frac{1}{2}(T^aT^b+T^bT^a), \\
:T^{a_1}\dots T^{a_l}:&=&\frac{1}{l!}\sum_{\sigma\in S_l}(T^{\sigma(a_1)}\dots
T^{\sigma(a_l)}).\nonumber
\end{eqnarray}
Any formal product of the generators can be expressed
in the above basis by using the commutation relations
(\ref{g3}), e.g.
\begin{equation}
\label{g7}
T^aT^b=\frac{1}{2}\{T^a,T^b\}+\frac{1}{2}[T^a,T^b]= :T^aT^b:+\frac{i}{2}f^{ab}_{\hspace{3mm} c} :T^c:.
\end{equation}

The new concept is to allow gauge transformations that are
enveloping algebra-valued:
\begin{equation}
\label{g8}
\hat{\La}_{\{\al\}}(\hat{x})=\sum_{l=1}^{\infty}\sum_{\textrm{\scriptsize{basis}}}\al_{a_1..
  a_l}^l(\hat{x}):T^{a_1}.. T^{a_l}:= \al_{a}(\hat{x}):T^a: +
  \al_{a_1 a_2}^2 :T^{a_1}T^{a_2}:+\dots \hspace{2mm}.
\end{equation}
With this definition we write this generalized transformation law
as follows:
\begin{equation}
\label{g9}
\de_{\{\al\}}\hat{\psi}(\hat{x}) =i\hat{\La}_{\{\al\}}(\hat{x}) \hat{\psi}(\hat{x}),
\end{equation}
where $\{\al\}$ denotes the set of the coefficient functions.
It is clear that the commutator of  two  enveloping
algebra-valued transformations will be
enveloping algebra-valued again. The price we have to pay
are the infinitely many parameters $\{\alpha\}$.
This is too expensive. But in the next chapter
we will get a price reduction. We will find in the next section that
we may define the enveloping algebra-valued transformation such 
that there will only be as many
independent parameters as there are in the Lie
algebra-valued case. Therefore it is worthwhile
to pursue this idea.

It can be seen that under these generalized gauge transformations a
covariant derivative
\begin{equation}
\label{g11}
\de_{\{\al\}}\Big(\hat{\cal{D}}_\mu\hat{\psi}(\hat{x})\Big)
=i\hat{\La}_{\{\al\}}(\hat{x})\hat{\cal{D}}_\mu \hat{\psi}(\hat{x})
\end{equation}
has to become enveloping algebra-valued as well, by adding an enveloping
algebra-valued gauge field:
\begin{eqnarray}
\label{g10}
\hat{\cal{D}}_\mu & =& \hat{D}_\mu - i
\hat{V}_{\mu},\nonumber\\
\hat{V}_{\mu}&=&\sum_{l=1}^{\infty}\sum_{\textrm{\scriptsize{basis}}}V^{l}_{\mu \hspace{0.5mm}a_1\dots
  a_l}:T^{a_1}\dots T^{a_l}:\hspace{1mm}.
\end{eqnarray}
Comparing with (\ref{g6a}), the gauge field $\hat{V}_{\mu}$ has to be
a vector field under $\kappa
$-Lorentz transformations. Therefore
each gauge field $\hat{V}^{l}_{\mu \hspace{0.5mm} a_1...a_l}$
has to transform vectorlike to guarantee
(\ref{g5a}).

A new feature arises due to the deformed coproducts (\ref{k12}) of the Dirac
operator which we used to define the covariant derivative.
We write (\ref{g11}) more explicitly
\begin{eqnarray}
\label{g12}
\de_{\{\al\}}\Big((\hat{D}_\mu - i
\hat{V}_{\mu})\hat{\psi}(\hat{x})\Big)&
=&i\hat{D}_\mu\Big(\hat{\La}_{\{\al\}}(\hat{x})
\hat{\psi}(\hat{x})\Big)+\hat{V}_{\mu}\hat{\La}_{\{\al\}}\hat{\psi}(\hat{x})-i\Big(\de_{\{\al\}}\hat{V}_{\mu}\Big)\hat{\psi}(\hat{x}) \nonumber\\
&\shouldid&i\hat{\La}_{\{\al\}}(\hat{x})(\hat{D}_\mu - i
\hat{V}_{\mu}) \hat{\psi}(\hat{x}).
\end{eqnarray}
Both $\hat{D}_n$ and $\hat{D}_i$ act in a non-trivial way on products
of functions. For example, (\ref{g11}) will be satisfied for $\hat{\cal{D}}_i$ if:
\begin{equation}
\label{g13}
(\de_{\{\al\}}\hat{V}_i)\hat{\psi}=(\hat{D}_i \hat{\La}_{\{\al\}})e^{-ia\hp_n}\hat{\psi}-i\lb\hat{V}_i,  \hat{\La}_{\{\al\}}\rb\hat{\psi}.
\end{equation}
If we want to use this formula such that it is independent of
$\hat{\psi}$, we see from (\ref{g13}) that the
gauge field has to be derivative-valued. Only then the transformation
\begin{equation}
\label{g14}
\de_{\{\al\}}\hat{V}_i=(\hat{D}_i \hat{\La}_{\{\al\}})e^{-ia\hp_n}-i\lb\hat{V}_i,  \hat{\La}_{\{\al\}}\rb,
\end{equation}
will lead to (\ref{g11}).
For $\hat{V}_n$ we can proceed in the
same way and find
\begin{eqnarray}
\label{g15}
\de_{\{\al\}}\hat{V}_n&=&(\hat{D}_n \hat{\La}_{\{\al\}})e^{-ia\hp_n}+
\Big((e^{ia\hp_n}-1) \hat{\La}_{\{\al\}}\Big)\hat{D}_n \nonumber\\
&&-ia(\hat{D}_je^{ia\hp_n}\hat{\La}_{\{\al\}})\hat{D}^j -i\lb\hat{V}_n,  \hat{\La}_{\{\al\}}\rb.
\end{eqnarray}
The gauge fields have to be derivative-valued to accommodate the first three terms on the
right hand side of equation (\ref{g15}) (first term on the right hand side of equation (\ref{g14})).
That the gauge fields appear as derivative-valued is a new feature
and is a direct consequence of the coproduct rules. We will discuss
more details in the next section.

The commutator of two covariant derivatives defines
a covariantly transforming object:
\begin{equation}
\label{g16}
\hat{\cal{F}}_{\mu\nu} = i \lb\hat{\cal{D}}_\mu ,\hat{\cal{D}}_\nu\rb.
\end{equation}
It will be enveloping algebra-  and
derivative-valued as it is the case for the gauge field. Its
transformation properties are tensorlike:
\begin{equation}
\label{g17}
\de_{\{\al\}}\hat{\cal{F}}_{\mu\nu}\hat{\psi} =
(\de_{\{\al\}}\hat{\cal{F}}_{\mu\nu})\hat{\psi}+ \hat{\cal{F}}_{\mu\nu}\de_{\{\al\}}\hat{\psi}.
\end{equation}
From the definition of the covariant derivative
follows:
\begin{equation}
\label{g18}
\de_{\{\al\}}\hat{\cal{F}}_{\mu\nu}\hat{\psi}
=i\hat{\La}_{\{\al\}}\hat{\cal{F}}_{\mu\nu}\hat{\psi}
=i(\hat{\La}_{\{\al\}}\hat{\cal{F}}_{\mu\nu}
-\hat{\cal{F}}_{\mu\nu}\hat{\La}_{\{\al\}})\hat{\psi}+ i\hat{\cal{F}}_{\mu\nu}\hat{\La}_{\{\al\}}\hat{\psi} .
\end{equation}
Comparing this with (\ref{g17})
and (\ref{g9}) shows the covariant transformation property
of $\hat{\cal{F}}_{\mu\nu}$:
\begin{equation}
\label{g19}
\de_{\{\al\}}\hat{\cal{F}}_{\mu\nu}
=i\lb \hat{\La}_{\{\al\}},\hat{\cal{F}}_{\mu\nu}\rb.
\end{equation}
The tensor ${\hat{\cal{F}}}_{\mu \nu}$ is derivative-valued.
Instead of expanding it in terms of the derivatives
$\hat{\partial}_\mu$, we can expand it in terms of
covariant derivatives $\hat{\cal{D}}_\mu$.

First we express $\hat{\partial}_n$ by $\hat{D}_\mu$ \cite{f1}:
\begin{equation}
e^{-ia\hp _n}=-ia{\hat D}_n+\sqrt{1+a^2{\hat D}_\mu{\hat D}^\mu}.\label{g20}
\end{equation}
Next we
replace  $\hat{D}_\mu$ by $\hat{\cal{D}}_\mu$ and subtract the additional
terms introduced that way :
\begin{equation}
\hat{D}_n=\hat{\cal{D}}_n+i\hat{V}_n. \label{g21}
\end{equation}
Each $\hat{V}_n$ will be derivative-valued again but
each derivative carries a factor $a$ and thus
contributes to the next  order in $a$. To first order
in $a$ we obtain from (\ref{g20}) and (\ref{g21}):
\begin{eqnarray}
e^{-ia\hp _n}&\rightarrow & 1-ia\hp _n =1-ia\hat{D}_n \nonumber\\
&=& 1-ia\hat{\cal{D}}_n+a\hat{V}_n . 
\end{eqnarray}
To lowest order in $a$ (compare with (\ref{g15})), $\hat{V}_n$ is not derivative-valued
and contributes  to
the term in $\hat{\cal{F}}_{\mu \nu}$ that has no derivatives.
Finally we arrive at an expression:
\begin{equation}
\hat{\cal{F}}_{\mu \nu}=\hat{F}_{\mu\nu}+\hat{T}^\rho_{\mu\nu}\hat{\cal{D}}_\rho+ \dots +
\hat{T}^{\rho _1\dots\rho _l}_{\mu\nu}:\hat{\cal{D}}_{\rho _1}\dots \hat{\cal{D}}_{\rho _l}:+\dots \ .\label{g22}
\end{equation}
The colons denote a basis in the free algebra of  covariant derivatives.
To each finite order in $a$ this expansion will have
a finite number of terms. The individiual terms
will transform like tensors as well:
\begin{eqnarray}
\hat{\cal{F}}_{\mu\nu}&\rightarrow & i\lb \hat{\La}_{\{\al\}},\hat{\cal{F}}_{\mu \nu}\rb
= i\lb \hat{\La}_{\{\al\}},\hat{F}_{\mu\nu} \rb
+ i\lb \hat{\La}_{\{\al\}},\hat{T}^\rho_{\mu\nu}\hat{\cal{D}}_\rho \rb
 \nonumber\\&&+ \dots + i\lb \hat{\La}_{\{\al\}},\hat{T}^{\rho _1\dots\rho_l}_{\mu\nu}
:\hat{\cal{D}}_{\rho_1}\dots \hat{\cal{D}}_{\rho_l}:\rb +\dots \label{g23}
\end{eqnarray}

When we apply this to a field $\hat{\psi}$ we find,
as before in (\ref{g17}) to (\ref{g19})
\begin{eqnarray}
\de_{\{\al\}}\hat{F}_{\mu\nu}
&=& i\lb \hat{\La}_{\{\al\}},\hat{F}_{\mu\nu}\rb, \label{g24}\\
\de_{\{\al\}}\hat{T}^{\rho_1\dots\rho_l}_{\mu\nu}
&=& i\lb \hat{\La}_{\{\al\}},\hat{T}^{\rho_1\dots\rho_l}_{\mu\nu}\rb. \label{g25}
\end{eqnarray}

Thus, $\hat{\cal{F}}_{\mu \nu}$ can be expanded in terms of a full series of
derivative-independent tensors. For the dynamics (Lagrangian) we are only
going to use the curvature-like term $\hat{F}_{\mu \nu}$.
It transforms like a tensor and reduces to the usual
field strength $F^0_{\mu\nu}$ for $a\rightarrow 0$. To first order we
get one torsion-like contribution $\hat{T}^\rho_{\mu \nu}$.

\section{Seiberg-Witten map}

In the previous chapter we saw that an enveloping
algebra-valued gauge transformation depends on an infinite
set of parameters. The same is true for the enveloping
algebra-valued gauge field, it depends on an infinite
set of vector fields. This gauge theory would feature an
infinite number of independent degrees of freedom.

This unphysical situation can be avoided if
we make the additional assumption that the transformation parameters $\La_{\{\al\}}$
depend on the usual, Lie algebra-valued gauge field
$A^0_{\mu a}$ \cite{SW}, \cite{madore}. We will
find that this dependence reduces the infinite number
of degrees of freedom of the deformed gauge theory to the finite number of
degrees of freedom of the Lie algebra gauge theory.

To find this dependence, known as the Seiberg-Witten map, we start from the
gauge transformation:
\begin{equation}
\label{s5}
\de_{\{\al\}} \hat{\psi} = i \hat{\La}_{\{\al\}} \hat{\psi}.
\end{equation}
The condition that this is actually a gauge transformation reads
\begin{equation}
\label{s6}
(\de_{\{\al\}} \de_{\{\be\}} -\de_{\{\be\}}\de_{\{\al\}})\hat{\psi} = \de_{\{\al\times\be\}}\hat{\psi}.
\end{equation}
We now introduce $\La_\al$ as opposed to $\La_{\{\al\}}$ referring
to solutions of the Seiberg-Witten map. We have to replace all
parameters in (\ref{g8}) by
\begin{equation}
\label{s4}
\al_{a_1 \dots a_l}^l (\hat{x}) \longrightarrow \al_{a_1 \dots a_l}^l(x; \al_a(x),
A^0_{\mu a}(x)).
\end{equation}
The parameters are functions of $x$, the
parameters $\alpha_a(x)\equiv\alpha_{a}^1(x)$
and  the gauge field $A^0_{\mu a}(x)$ as well as of their derivatives.
Since we define that the noncommutative gauge parameters have a
functional dependence only on commuting variables, we have to use the $\star$-product
formalism. We choose as a
starting point \cite{moeller}
\begin{equation}
\label{s4a}
\de_\al \psi = i \La_\al \star \psi\quad\textrm{with}\quad(\de_{\al} \de_{\be} -\de_{\be}\de_{\al})\psi = \de_{\al\times\be}\psi.
\end{equation}

The Lie algebra-valued gauge field
$A^0_\mu$ (in the following we omit all explicit dependence
on coordinates $x$):
\begin{equation}
\label{s1}
A^0_\mu=A^0_\mu(x)=\sum_a A^0_{\mu a}(x)T^a
\end{equation}
has the transformation property
\begin{equation}
\label{s2}
\de_\al A^0_\mu=\p_\mu \al -i[ A^0_\mu, \al] ,
\end{equation}
where $\alpha=\al_a(x)T^a$ is Lie algebra-valued as well.

The gauge parameter $\La_\al$ depends on $A^0_\mu$ and because of
(\ref{s2}) $\delta_\alpha\Lambda_\beta$ is not zero.
We take this into account when we write (\ref{s4a}) more explicitly
\begin{equation}
\label{s7}
(\de_\al\de_\be -\de_\be\de_\al)\psi = (\La_\al\star\La_\be
-\La_\be\star\La_\al)\star\psi +i(\de_\al\La_\be -\de_\be\La_\al)\star\psi=\de_{\al\times\be} \psi \, .
\end{equation}
 That (\ref{s7}) has a solution can
be seen on more general grounds \cite{kontsevich} (also \cite{zumino},
\cite{quadri} and \cite{brandt}). Here we construct
a solution by a power series expansion in the
deformation parameter $a$:
\begin{equation}
\label{s8}
\La_\al = \al+a \La_\al^1+ \dots + a^k \La_\al^k +\dots \ .
\end{equation}

In this paper we will consider only the first order
term in $a$ to make the formalism transparent. Assuming $a$ to be small, only
the
leading term will be of relevance for phenomenological
applications. We have however calculated all quantities also to second
order and have checked the validity of the statements made here.

We expand
(\ref{s7}) to first order in $a$:
\begin{eqnarray}
\label{s9}
&&\La^0_\al\La^1_\be+ \La^1_\al\La^0_\be+ \La^0_\al\star\La^0_\be|_{{\mathcal O}(a)}
-\La^0_\be\La^1_\al\\
&&-\La^1_\be\La^0_\al-\La^0_\be\star\La^0_\al|_{{\mathcal
    O}(a)} +i(\de_\al\La^1_\be
-\de_\be\La^1_\al)=i\La^1_{\al\times\be} \,\nonumber
\end{eqnarray}
or, using $\Lambda^0_\alpha = \alpha, \Lambda^0_\beta = \beta$ and the
explicit form of the $\star$-product,
\begin{equation}
\label{s9a}
\lb\al,\La^1_\be\rb + \lb\La^1_\al,\be\rb +\frac{i}{2}x^\la
C^{\mu\nu}_\la \{\partial_\mu \al, \partial_\nu
\be\}+i(\de_\al\La^1_\be
-\de_\be\La^1_\al)=i\La^1_{\al\times\be} \, .
\end{equation}
To first order in $a$ the noncommutative structure contributes a term
from the $\star$-product, which forbids
$\Lambda^1_\alpha$ equal zero.
The equation (\ref{s9a}) is an
inhomogeneous linear equation in $\Lambda^1_\al$, with the solution:
\begin{equation}
\label{s10}
\La^1_{\al}=-\frac{1}{4}x^\la C^{\mu\nu}_\la \{A^0_\mu, \partial_\nu
\al\},
\end{equation}
where $C^{\mu\nu}_\lambda$ are the structure
constants of the coordinate algebra. More explictly this is
\begin{equation}
\Lambda^1_\al ={a\over 4}x^j\left( \{A^0_j,\partial _n\alpha\} -\{A^0_n,\partial _j\alpha\} \right) .
\label{s14}
\end{equation}
This solution is hermitean for real fields
$A^0_{\mu a}(x)$ and real parameters $\alpha_a(x)$.
That this specific solution of the inhomogeneous equation is not unique and that it is possible to add to it
 solutions of the
homogeneous equation
\begin{equation}
\label{s11}
\lb\al,\La^1_\be\rb+ \lb\La^1_\al,\be\rb
 +i(\de_\al\La^1_\be -\de_\be\La^1_\al)=i\La^1_{\al\times\be} \,
\end{equation}
has been discussed thoroughly in many places (e.g. \cite{moeller}, \cite{brandt}). There are no new aspects
to this question in first order in $a$ in this particular model.

With a solution for $\La_\al^1$ at our disposition, it is possible to express a "matter"
field  $\psi$ (i.e. field in the fundamental representation) in terms of $A^0_\mu$ and a matter
field $\psi^0$ of the standard gauge theory
\begin{equation}
\delta_\alpha \psi ^0=i\alpha\psi ^0. \label{s24}
\end{equation}

Up to first order in $a$, (\ref{s4a}) is solved
by:
\begin{equation}
\psi =\psi^0 -{1\over 2}x^\la C^{\mu\nu}_\la A^0_\mu\partial _\nu\psi ^0
+{i\over 8}x^\la C^{\mu\nu}_\la\lb A^0_\mu,A^0_\nu\rb \psi ^0 \label{s25}.
\end{equation}

The same way as we express $\psi$ in terms of
$A^0_\mu$ and $\psi^0$, we can define the Seiberg-Witten map for
gauge fields (they are in the adjoint representation of the enveloping algebra).
When we derived the respective formulae in the previous section, we
discovered that the gauge fields have to be derivative-valued.
Therefore we have to discuss solutions of the Seiberg-Witten map for
the following relations:
\begin{equation}
\label{s11a}
\de_{\al}V_i=(D^*_i \La_{\al})e^{-ia\partial_n}-i\lb V_i\ds \La_{\al}\rb,
\end{equation}
and
\begin{equation}
\label{s11b}
\de_{\al}V_n=(D^*_n \La_{\al})e^{-ia\partial_n}+
\Big((e^{ia\partial_n}-1) \La_{\al}\Big)D^*_n-ia(D^*_je^{ia\partial_n}\La_{\al})D^{*j} -i\lb V_n\ds  \La_{\al}\rb.
\end{equation}
It is technically not simple to solve these two equations (especially
since the second is a sum of several terms with different dependence on
 derivatives), but conceptually there are no further problems.
Without going into details we present the solution up to first order
in $a$:
\begin{eqnarray}
V_i &=& A^0_i-iaA^0_i\partial _n -{ia\over 2}\partial _n A^0_i-{a\over
  4}\{A^0_n,A^0_i\}\nonumber+{1\over 4}x^\la C^{\mu\nu}_\la \big( \{ F^0_{\mu i},A^0_\nu\} -\{ A^0_\mu ,\partial _\nu A^0_i\}\big),
\label{s15} \\
V_n &=& A^0_n-iaA^{0j}\partial _j -{ia\over 2}\partial _jA^{0j}-{a\over 2}A^0_jA^{0j}+{x^\la\over 4} C^{\mu\nu}_\la \big( \{ F^0_{\mu n},A^0_\nu\}
-\{ A^0_\mu ,\partial _\nu A^0_n\}\big) .\label{s16}
\end{eqnarray}
Here $F^0_{\mu \nu}=\partial _\mu A_\nu^0 - \partial _\nu A^0_\mu -i\lb A^0_\mu, A^0_\nu\rb $ is the
field strength of the undeformed gauge theory.
We emphasize the
dependence on derivatives in the
terms $A^0_i\partial_n$ and $A^{0j}\partial _j$.

From the covariant derivative ${\cal D}_\mu= D^*_\mu-iV_\mu$ we can calculate
${\cal F}_{\mu \nu}=i\lb\cal{D}_\mu\ds \cal{D}_\nu\rb $ to first order in $a$. As discussed in the
previous section, it will
be of first order in the derivatives, the sum of a curvature-like term
and a torsion-like term:
\begin{equation}
{\cal{F}}_{\mu \nu}=F_{\mu\nu}+T^\rho_{\mu\nu}{\cal{D}}_\rho .\label{s17}
\end{equation}

The result is (up to first order in $a$):
\begin{eqnarray}\label{s18}
F_{ij} &=& F_{ij}^0 -ia{\cal D}_n^0F_{ij}^0
+{x^\la\over 4} C^{\mu\nu}_\la \big( 2\{F^0_{\mu i},F_{\nu j}^0\}+\{ {\cal D}^0_\mu F_{ij}^0,A^0_\nu \} -\{ A^0_\mu,\partial _\nu F_{ij}^0 \}\big) , \\
T_{ij}^\mu &=& -2ia\eta ^\mu _{\ n}F_{ij}^0, \label{s19}\\
F_{nj} &=& F_{nj}^0 -{ia\over 2}{\cal D}^{\mu 0}F_{\mu j}^0+{x^\la\over 4} C^{\mu\nu}_\la \big( 2\{F^0_{\mu n},F_{\nu j}^0\}
+\{ {\cal D}^0_\mu F_{nj}^0,A^0_\nu \} -\{ A^0_\mu,\partial _\nu F_{nj}^0 \}\big) , \label{s20}\\
T_{nj}^\mu &=& -ia\eta ^{\mu l}F_{lj}^0-ia\eta ^\mu _{\ n}F^0_{nj}. \label{s21}
\end{eqnarray}
These quantities transform covariantly
\begin{eqnarray}
\de_\alpha F_{\mu\nu}
&=& i\lb \Lambda_\alpha \ds F_{\mu\nu}\rb, \label{s22}\\
\de_\alpha T^\rho _{\mu\nu}
&=& i\lb \Lambda _\alpha \ds T^\rho _{\mu\nu}\rb. \label{s23}
\end{eqnarray}
Now we have all the ingredients to construct to first
order in $a$ a gauge theory based on the noncommutative
spaces defined by (\ref{k1}) in terms of the usual fields $A^0_\mu$
and $\psi^0$.

The dynamics of the gauge field can be formulated with the tensor
$F^{\mu \nu}$
\begin{equation}
{\cal L}_{\textrm{\tiny{gauge}}}=c\ \textrm{Tr}\hspace{0.8mm} \big(F^{\mu \nu}\star F_{\mu \nu}\big). \label{s26}
\end{equation}
Note, however, that $\textrm{Tr}\hspace{0.8mm}\big(F_{\mu \nu}\star F^{\mu
  \nu}\big)$ is not invariant because the coordinates do not commute. The Lagrangian
${\cal L}_{\textrm{\tiny{gauge}}}$ will render an action gauge
  invariant, if it is formulated with an integral
with the trace property\footnote{To attain the trace
  property, a measure function can be introduced (compare
  \cite{f1}). Since the measure function does in general not vanish in
  the limit $a\rightarrow 0$, it should be compensated without
  spoiling the gauge invariance of the action. This is possible,
  leading however to additional first-order terms in the action
  (compare e.g. \cite{steinac}).}. The trace will also
depend on the representation of the generators
$T^a$ because higher products of the generators will
enter through the enveloping algebra (for a detailed discussion of
  this issue, see \cite{aschieri}).

To first order in $a$, when written in terms of $A^0_{\mu}$,
we obtain the following expression for the gauge part of the Lagrangian
(choosing in analogy to the undeformed theory $c=-\frac{1}{4}$)
\begin{eqnarray}
\label{s26a}
{\cal L}_{\textrm{\tiny{gauge}}}|_{\mathcal{O}(a)}&=&-\frac{i}{8} x^\la
 C^{\rho\sigma}_\la \textrm{Tr}\hspace{0.8mm}\Big( \mathcal{D}^0_\rho F^{0\mu
 \nu}\mathcal{D}^0_\sigma F^0_{\mu \nu}+\frac{i}{2} \{A^0_\rho,
 (\partial_\sigma +\mathcal{D}^0_\sigma)(F^{0\mu \nu}F^0_{\mu \nu})\}\\
 && -i \{F^{0\mu\nu},\{F^0_{\mu\rho},F^0_{\nu\sigma}\}\}\Big)+\frac{ia}{4}  \textrm{Tr}\hspace{0.8mm}\Big( \mathcal{D}^0_n (F^{0\mu \nu} F^0_{\mu \nu})-\{\mathcal{D}^0_\mu F^{0\mu j}, F^0_{nj}\}\Big), \nonumber
\end{eqnarray}
where $\mathcal{D}^0_\mu=\partial_\mu -iA^0_\mu$ (or adjoint
$\mathcal{D}^0_\mu\cdot=\partial_\mu\cdot -i\lb A^0_\mu,\cdot\rb$
acting on $F^0_{\mu\nu}$). Cyclicity of the trace allows several
simplifications on the terms on the right-hand side.

The matter part of the theory will be the gauge covariant version of
the free Lagrangian as it was developed in \cite{f1}
\begin{equation}
{\cal L}_{\textrm{\tiny{matter}}}=\bar{\psi}\star \left( i\gamma ^\mu{\cal D}_\mu -m\right) \psi .\label{s27}
\end{equation}

To first order in $a$, when written in terms of $A^0_{\mu}$
and $\psi^0$, we obtain
\begin{eqnarray}
\label{s28}
{\cal L}_{\textrm{\tiny{matter}}}|_{\mathcal{O}(a)}&=&\frac{i}{2}x^\nu
C^{\rho\sigma}_\nu\overline{\mathcal{D}^0_\rho\psi^0}\
\mathcal{D}^0_\sigma \left( i\gamma ^\mu{\cal D}^0_\mu -m\right)
\psi^0-\frac{i}{2}x^\nu C^{\rho\sigma}_\nu \bar{\psi^0}\gamma ^\mu
F^0_{\mu\rho}\mathcal{D}^0_\sigma\psi^0\nonumber \\
&&+\frac{a}{2} \bar{\psi^0}\gamma ^j\mathcal{D}^0_n\mathcal{D}^0_j\psi^0 + \frac{a}{2} \bar{\psi^0}\gamma ^n\mathcal{D}^0_j\mathcal{D}^{0j}\psi^0.
\end{eqnarray}

These are the Lagrangians which define the dynamics
on the $\kappa$-deformed Minkowski space.

\section{Gauge transformations and the $\kappa$-Lorentz algebra}

Our concept of gauge transformations on noncommutative spaces rests on the Seiberg-Witten
map. With the help of this map gauge transformations can be realized
in the enveloping algebra of the Lie algebra
\begin{eqnarray}
\label{t1}
\phi'=\phi+\de_\al\phi& =&\phi+ i\La_\al \star \phi\nonumber\\
\textrm{with}\qquad (\de_\al\de_\be-\de_\be\de_\al) \phi &=& \de_{\al\times \be}\phi.
\end{eqnarray}
To find such a realization it turned out to be necessary that $\La_\al$ depends on
the standard Lie algebra-valued gauge field $A^0_\mu$ and its
derivatives. Therefore under a gauge transformation $\La_\al$ will transform
as well and (\ref{t1}) leads to
\begin{equation}
\label{t3}
i(\de_\al \La_\be-\de_\be\La_\al)\star \phi +(\La_\al\star\La_\be-\La_\be\star\La_\al)\star\phi = i\La_{\al\times \be}\star\phi.
\end{equation}
In this section we want to see how these equations behave under the
$\kappa$-deformed Lorentz transformations. Only $M^{*in}$ has a deformed coproduct
rule (compare with (\ref{k21})):
\begin{equation}
\label{t4}
M^{*in} \left( f\star g\right) = \left( M^{*in} f \right) \star g
+ \left(e^{ia \pat_n^*} f\right) \star \left( M^{*in} g\right) +ia \left( \pat_j^* f \right) \star \left( M^{*ij} g\right).
\end{equation}
and therefore we will restrict our discussion to $N_\epsilon=
\epsilon_i M^{in}$.

A scalar  field  transforms as follows:
\begin{equation}
\label{t6}
\tilde{\phi} = \phi -N_\epsilon^* \phi,
\end{equation}
where $N_\epsilon^*$ acts on the coordinates, compare with (\ref{g2a}).
This transformation can be inverted, to first order in
$\epsilon$:
\begin{equation}
\label{t7}
\phi= \tilde{\phi} + N_\epsilon^* \phi .
\end{equation}
We assume that $\La_\al$ transforms like a scalar field.

First we compute $\widetilde{\phi'}$, by applying (\ref{t6}) to
(\ref{t1}):
\begin{equation}
\label{t8}
\widetilde{\phi'}= \phi +i\La_\al\star \phi - (N_\epsilon^*
\phi)-iN_\epsilon^*(\La_\al\star \phi).
\end{equation}
For evaluating the last term in (\ref{t8}), the coproduct (\ref{t4}) has to be used.

Next we compute $\tilde{\phi}\hspace{1mm}'$ by applying (\ref{t1}) to (\ref{t6}):
\begin{equation}
\label{t9}
\tilde{\phi}\hspace{1mm}'= \phi- (N_\epsilon^*
\phi) +i\La_\al\star \phi -iN_\epsilon^*(\La_\al\star \phi).
\end{equation}
This shows that the two transformations commute.
When we use (\ref{t7}), the gauge transformation (\ref{t8}) can be
written as a gauge transformation on $\tilde{\phi}$:
\begin{equation}
\label{t10}
\de_\al\tilde{\phi}= i\La_\al\star \tilde{\phi} +i\La_\al\star (N_\epsilon^*\tilde{\phi})
-iN_\epsilon^*(\La_\al\star \tilde{\phi}).
\end{equation}
We draw the commuting diagram to illustrate the result
\begin{eqnarray}
\label{t11}
\phi&\stackrel{\al}{\longrightarrow}& \phi'\nonumber\\
\epsilon\downarrow&&\downarrow\epsilon\\
\tilde{\phi}&\stackrel{\al}{\longrightarrow}&\widetilde{\phi'}\equiv\tilde{\phi}\hspace{1mm}'\hspace{2mm}.\nonumber
\label{diagramm}
\end{eqnarray}

The gauge transformations on $\tilde{\phi}$ - the $\kappa$-Lorentz
transformed scalar field - is now defined by (\ref{t10}).
It remains to be shown that (\ref{t10}) realizes the gauge group as well:
\begin{equation}
\label{t12}
(\de_\al\de_\be-\de_\be\de_\al)\tilde{\phi}=\de_{\al\times\be} \tilde{\phi}.
\end{equation}
It is easier to compute $\de_\be\de_\al\tilde{\phi}$ from (\ref{t9})
and to use (\ref{t1}). We make use of (\ref{t4}) and after some
rearrangements
we obtain:
\begin{eqnarray}
\label{t13}
(\de_\be\de_\al-\de_\al\de_\be)\tilde{\phi}
&=& \Big(i(\de_\be\La_\al -
\de_\al\La_\be) - (\La_\al\star\La_\be -
\La_\be\star\La_\al)\Big)\star\phi \nonumber\\
&&-N^*_\epsilon\Big(i(\de_\be\La_\al -
\de_\al\La_\be) - (\La_\al\star\La_\be -
\La_\be\star\La_\al)\Big)\star\phi\\
&&-e^{ia\partial^*_n}\Big(i(\de_\be\La_\al -
\de_\al\La_\be) - (\La_\al\star\La_\be -
\La_\be\star\La_\al)\Big)\star N^*_\epsilon\phi\nonumber\\
&&+ia\partial^*_j\Big(i(\de_\be\La_\al -
\de_\al\La_\be) - (\La_\al\star\La_\be -
\La_\be\star\La_\al)\Big)\star\epsilon_l M^{lj}\phi.\nonumber
\end{eqnarray}
We use the condition (\ref{t3}) again and obtain the result
(\ref{t12}). This demonstrates that (\ref{t10}) is a gauge
transformation.

It is also possible to verify the result (\ref{diagramm}) by a direct calculation.
We start with the solution of the Seiberg-Witten map (\ref{s14})
\begin{eqnarray}
\Lambda_{\alpha} &=&\alpha -\frac{1}{4}x^\la C^{\mu\nu}_\la \{A^0_\mu , \partial_\nu \alpha \} +\mathcal{O}(a^2)   \nonumber \\
 &=:& \alpha + \Lambda_{\alpha}^{1} + \mathcal{O}(a^2)
\end{eqnarray}
and (\ref{s25})
\begin{eqnarray} \label{SW-sol_phi}
\phi &=& \phi^0-\frac{1}{2} x^{\mu}C_{\mu}^{\rho \sigma}
A^0_{\rho}\partial_{\sigma}\phi^0+\frac{i}{8}
x^{\mu}C_{\mu}^{\rho\sigma} [A^0_{\rho},A^0_{\sigma}]\phi^0 + \mathcal{O}(a^2)  \nonumber \\
&=:& \phi^{0}+\phi^{1} +\mathcal{O}(a^2) \, .
\end{eqnarray}

We first apply $M^{*in}$ to (\ref{SW-sol_phi}) and gauge-transform the undeformed fields afterwards.
This has to be equal to $M^{*in}$ applied to $\delta_{\alpha}\phi=i\Lambda_{\alpha}\star\phi$ up to first order in
$a$. Applying $M^{*in}$ on $\delta_{\alpha}\phi$, the coproduct (\ref{t4}) has to be taken into
account and we obtain
\begin{equation} \label{N_on_delta-phi}
M^{*in}(\Lambda_\alpha\star\phi)=
(M^{*in}\Lambda_\alpha)\star\phi+(e^{ia\partial^*_n}\Lambda_\alpha)\star(M^{*in}\phi)
 +\, ia(\partial^*_j
\Lambda_\alpha)\star(M^{ij*}\phi) \, .
\end{equation}
To write this explicitly to first order we need the operators (\ref{k21}) expanded up to first order in $a$:
\begin{equation} \label{N_fist_order}
M^{*in}  =
x^i\partial^n-x^n\partial^i+\frac{ia}{2}x^i\partial_{\mu}\partial^{\mu}
-\frac{ia}{2}x^{\mu}\partial_{\mu}\partial^{i}
 =:  M^{*in}_{0} + M^{*in}_{1}  \,
\end{equation}
and
\begin{eqnarray}
M^{ij*}=x^i\partial^j-x^j\partial^i =:  M^{ij*}_{0} \,.
\end{eqnarray}
Now we obtain from (\ref{N_on_delta-phi}):
\begin{eqnarray}
i M^{*in} (\Lambda_\alpha\star\phi)|_{\mathcal{O}(a)}
&=& i(M^{*in}_{1}\alpha)\phi^{0}+i\alpha (M^{*in}_{1} \phi^{0}) -
 \frac{1}{2}x^{\mu}C_{\mu}^{\rho\sigma}\partial_{\rho}(M^{*in}_{0}\alpha)\partial_{\sigma}\phi^{0}
\nonumber \\
&& -\,\frac{1}{2}x^{\mu}C_{\mu}^{\rho\sigma}\partial_{\rho}\alpha\partial_{\sigma}(M^{*in}_{0}\phi^{0})
 +i(M^{*in}_{0}\alpha)\phi^{1} \\
&& +\, i\alpha (M^{*in}_{0}\phi^{1}) +i(M^{*in}_{0}\Lambda_{\alpha}^{1})\phi^{0}
 + i\Lambda_{\alpha}^{1}(M^{*in}_{0}\phi^{0})\nonumber\\
&& -\, a\partial_n \alpha (M^{*in}_{0}\phi^{0})
-a\partial_j \alpha (M^{ij*}_{0} \phi^{0}) \, .\nonumber
\end{eqnarray}
Notice that
\begin{equation} \label{Vereinfachung1}
\delta_{\alpha}(M^{*in}_{0}\phi^{1})=iM^{*in}_{0}(\frac{i}{2}x^{\mu}C_{\mu}^{\rho\sigma}\partial_{\rho}\alpha\partial_{\sigma}\phi^{0}+\alpha\phi^{1}+\Lambda_{\alpha}^{1}\phi^{0})
\, ,
\end{equation}
since $\phi^{1}$ was constructed as solution for the
Seiberg-Witten map (\ref{s25}). Besides it can be shown by direct calculation that
\begin{eqnarray}
iM^{*in}_{1}(\alpha)\phi^{0}+i\alpha M^{*in}_{1}
(\phi^{0})
&=&iM^{*in}_{1} (\alpha\phi^{0})- ax^j
\partial_{\mu}\alpha\partial^{\mu}\phi^{0}\nonumber\\&&+\frac{a}{2}x^{\mu} \partial_{\mu}\alpha\partial^{u}\phi^{0}+\frac{a}{2}x^{\mu}\partial^{i}\alpha\partial_{\mu}\phi^{0}
\label{Vereinfachung2}
\end{eqnarray}
as well as
\begin{eqnarray}
&& -\frac{1}{2}x^{\mu}C_{\mu}^{\rho\sigma}\partial_{\rho}(M^{*in}_{0}\alpha)\partial_{\sigma}\phi^{0} -
\frac{1}{2}x^{\mu}C_{\mu}^{\rho\sigma}\partial_{\rho}\alpha\partial_{\sigma}(M^{*in}_{0}\phi^{0}) =
\nonumber\\
&& = -M^{*in}_{0}(\frac{1}{2}x^{\mu}C_{\mu}^{\rho\sigma}\partial_{\rho}\alpha\partial_{\sigma}\phi^{0})
+\frac{1}{2}M^{*in}_{0}(x^{\mu})C_{\mu}^{\rho\sigma}\partial_{\rho}\alpha \partial_{\sigma}\phi^{0}
\nonumber\\
&& -\frac{1}{2}x^{\mu}C_{\mu}^{\rho\sigma}\Big(
\partial_{\rho}(M^{*in}_{0})(\alpha)\partial_{\sigma}\phi^{0}
+\partial_{\rho}\alpha\partial_{\sigma}(M^{*in}_{0}\phi^{0})\Big) \, ,
\label{Vereinfachung3}
\end{eqnarray}
where $\partial_{\rho}(M^{*in}_{0})(\alpha) := \eta^i_\rho \partial^n\al-\eta^n_\rho \partial^i\al$.
Then equations (\ref{Vereinfachung1}), (\ref{Vereinfachung2}) and (\ref{Vereinfachung3}) yield
\begin{eqnarray}
 M^{*in}(\Lambda_\alpha\star\phi)|_{\mathcal{O}(a)}
&=&\delta_{\alpha}(M^{*in}_{1}\phi^{0})+\delta_{\alpha}(M^{*in}_{0}\phi^{1})-ax^i\partial_{\mu}\alpha\partial^{\mu}\phi^{0}
+\frac{a}{2} x^{\mu} \partial_{\mu}\alpha\partial^{i}\phi^{0}
\nonumber \\
&& +\frac{a}{2}x^{\mu}\partial^{i}\alpha\partial_{\mu}\phi^{0}
+\frac{1}{2}M^{*in}_{0}(x^{\mu})C_{\mu}^{\rho\sigma}\partial_{\rho}\alpha \partial_{\sigma}\phi^{0}
\label{Vereinfachung4}\\
&& -\frac{1}{2}x^{\mu}C_{\mu}^{\rho\sigma}\Big(
\partial_{\rho}(M^{*in}_{0})(\alpha)\partial_{\sigma}\phi^{0}
+\partial_{\rho}\alpha\partial_{\sigma}(M^{*in}_{0}\phi^{0})\Big) \, .\nonumber
\end{eqnarray}
Calculation shows that the last terms on the right hand side all cancel and we end up with
\begin{equation}
(M^{*in}\delta_{\alpha}\phi)|_{\mathcal{O}(a)} = \delta_{\alpha}(M^{*in}_{1}\phi^{0})+\delta_{\alpha}(M^{*in}_{0}\phi^{1}) \,.
\end{equation}
Hence we showed  explicitly up to first order in $a$ that (\ref{diagramm}) is true.

\section*{Acknowledgment}
M. D.  gratefully acknowledges the support of the Deutscher
Akademischer Austauschdienst.

We are very grateful to Larisa Jonke, Efrossini Tsouchnika and Michael  Wohlgenannt for intensive discussions and many very valuable remarks.

\end{document}